\documentclass[journal]{IEEEtran}
\usepackage{cite}

\ifCLASSINFOpdf
  \usepackage[pdftex]{graphicx}
  \graphicspath{Figures/}
  \DeclareGraphicsExtensions{.pdf,.jpeg,.png}
\else
\fi
\usepackage{amsmath}
\usepackage{amssymb}
\usepackage{url}

\usepackage{tikz}
\usepackage{textcomp}
\usepackage{hyperref}
\usepackage{lipsum}

\newcommand\copyrighttext{%
  \footnotesize This work has been submitted to the IEEE for possible publication. Copyright may be transferred without notice, after which this version may no longer be accessible.}
\newcommand\copyrightnotice{%
\begin{tikzpicture}[remember picture,overlay]
\node[anchor=south,yshift=10pt] at (current page.south) {\fbox{\parbox{\dimexpr\textwidth-\fboxsep-\fboxrule\relax}{\copyrighttext}}};
\end{tikzpicture}%
}

\begin{document}
%
\title{Cooling in poor air quality environments - \\Impact of fan operation on particle deposition}
%
%
%

\author{Jason Stafford
        and~Chen~Xu 
\thanks{J. Stafford is with the School of Engineering, University of Birmingham, Birmingham B15 2TT, United Kingdom. (e-mail: j.stafford@bham.ac.uk). }
\thanks{C. Xu is with Bell Labs - CTO, Nokia, Murray Hill, NJ 07974 USA.}
\thanks{Manuscript prepared February 17, 2021.}}

%
%

\markboth{\textit{Submitted for consideration} to IEEE Special issue in Trans. CPMT, February 2021}%
{Stafford and Xu: Cooling in poor air quality environments - Impact of fan operation on particle deposition}
%



\maketitle

\copyrightnotice

\begin{abstract}
Environmental pollutants are a source for reliability issues across data center and telecommunications equipment. A primary driver of this is the transport and deposition of particle matter (PM$_{2.5}$, PM$_{10}$) on printed circuit boards, electronic components and heat exchange surfaces. This process is enhanced by turbulent air flows generated from cooling fans. Particle pollutants can persist after contemporary filtering, highlighting the importance of elucidating particle transport mechanisms and utilising this information to design robust equipment. This study investigates particle transport behaviour arising from axial fans operating under varied aerodynamic conditions. Transient, multi-phase numerical simulations were performed to model the flow of millions of microscale particles in air and determine their fate. Across a comprehensive range of fan operation conditions, from aerodynamic stall to free delivery, non-dimensional deposition velocities spanned an order of magnitude. Deposition profiles vary from monotonic to non-monotonic behaviour, influenced by local flow impingement, blade tip vortices, and shear velocity. A simple flow control solution that mitigates the factors influencing deposition has been demonstrated for equipment already deployed. The findings and numerical methods can be applied for the optimization of fan-cooled equipment intended for indoor and outdoor environments where air quality is poor, or pollution levels are high.
\end{abstract}


%
\IEEEpeerreviewmaketitle

\section{Introduction}
%
%
%
%
\IEEEPARstart{T}he reliability challenge posed by atmospheric pollutants has grown significantly since it was highlighted over thirty years ago by Comizolli et al.~\cite{Comizzoli1986}. Corrosive gases (e.g. SO$_{2}$, H$_{2}$S, NO$_{x}$) and particle contaminants accelerate electronic equipment failures primarily through material corrosion and leakage current effects~\cite{Frankenthal1993}. Micron and sub-micron scale particles have diverse constituents, deposit onto small feature electronic devices, and persist globally. They originate naturally from wind-driven dispersion of land dust, sea salt, or from human-related activities and industrial emission sources~\cite{Xiao2018, Song2013, Yue2006}. These particles can be hygroscopic and contain soluble salts, providing the ingredients to deliquesce in environments with moderate relative humidity (typically 50-65\%~\cite{Frankenthal1993}). 

The sensitive interplay between temperature, relative humidity, particle matter (PM) and the resulting reliability issues has led to recommendations for allowable operating environments of information technology equipment~\cite{Shah2017}. There are, of course, many instances where controlling these environments is either not possible (e.g. outdoor equipment deployments) or disadvantageous on cooling efficiency. Free cooling of data centers using airside economizers reduces air-conditioning energy demands by bringing in outside air to cool equipment~\cite{Shah2017}. This introduces a reliability risk by exposing the electronic equipment to atmospheric pollutants and recent studies have begun to investigate reliability issues~\cite{Shah2019}. 

Identifying and excluding problematic contaminants would avoid these reliability concerns. Filtering is the widespread approach for removing large particles (e.g. $d_p >$ 10 $\mu$m), however, most fine particle matter remain. High-efficiency clean room filters are generally impractical (e.g. above MERV 13 rating), introducing energy penalties in forced-cooled equipment, as air movers must overcome the substantial flow resistance. The particles that do enter the equipment either pass through or deposit onto surfaces via a number of mechanisms, ranging from gravitational settling to inertia-driven forced flows~\cite{Tencer2008,Stafford2021}. 

\begin{figure}[!t]
\centering
\includegraphics[width=3.0in]{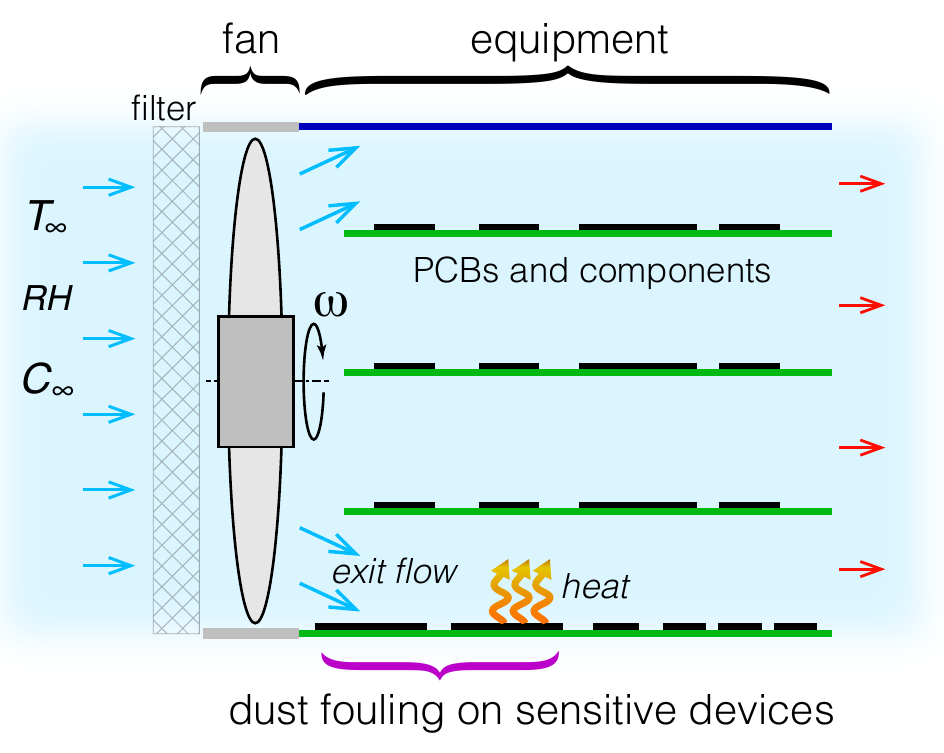}
\caption{A typical arrangement for forced cooling electronic equipment and the region susceptible to enhanced particle deposition and device failures.}
\label{figSketch}
\end{figure}

Inertia effects can dominate in fan-cooled systems. Preferential deposition patterns in electronic equipment have recently been studied using \textit{in situ} particle image velocimetry measurements and  observations from long-term installations in the field~\cite{Stafford2021}. These patterns of enhanced particle deposition were found to be a consequence of the fan exit flow angle and local turbulent events that enhance mass transport. Notably, the deposition patterns are non-unique, appearing in other fan-cooled systems and data center environments also (e.g. figure 1 from ref.~\cite{Stafford2021} and figure 3 from ref.~\cite{Shah2019}). These push flow configurations, illustrated in Fig. ~\ref{figSketch}, are regularly used for cooling electronic systems. Utilising the swirling turbulent exit flow can provide gains in local heat dissipation ~\cite{Grimes2001}.

Mass transfer is also enhanced by turbulent air transport, accelerating the formation of particle deposition sites which become a nucleus for corrosion and leakage current problems. Therefore, revealing the fundamental connection between deposition and the flow delivered by air movers is crucial for informing the design of robust technologies exposed to atmospheric pollution. This would also support the advancement of next-generation multi-phase predictive tools that enable package and equipment designers to reduce the potential risks to reliability. Indeed, this strategy aligns with the exemplary work and vision of Kraus and Bar-Cohen \cite{Kraus1983} and Bar-Cohen \cite{BarCohen1992}, who showed on many occasions how fundamental studies in thermal sciences can form the basis of developing physical design methodologies and solutions that address thermally-induced device failures.   

Previous experiments have shown that fan exit flows play a key role in mass transport of fine particles~\cite{Stafford2021}. The aim of this study is to investigate the influence of cooling fan operation point on the deposition of particle pollutants. Fan operation point is influenced by the airflow resistance of the equipment being cooled, filter installations, filtration efficiency changes over time, and other external effects such as backpressure resulting from aisle containment configurations~\cite{Khalili2019}. The present investigation has been conducted using a multi-phase numerical method that can be readily incorporated into existing computational fluid dynamics codes currently used in the thermal design of electronics. In addition, a simple flow control method has been presented that reduces the impact of the exit flow driving enhanced particle deposition. This was demonstrated by installing the solution in a fan-cooled telecommunications equipment.

\section{Methods}

\noindent The transport of air and atmospheric particles due to rotating fan flows was numerically modelled. This section describes the fan design, particle properties, and numerical techniques used to perform these multi-phase simulations.

\subsection{Fan characteristics and particle properties}

\begin{figure}[t]
\centering
\includegraphics[width=3.3in]{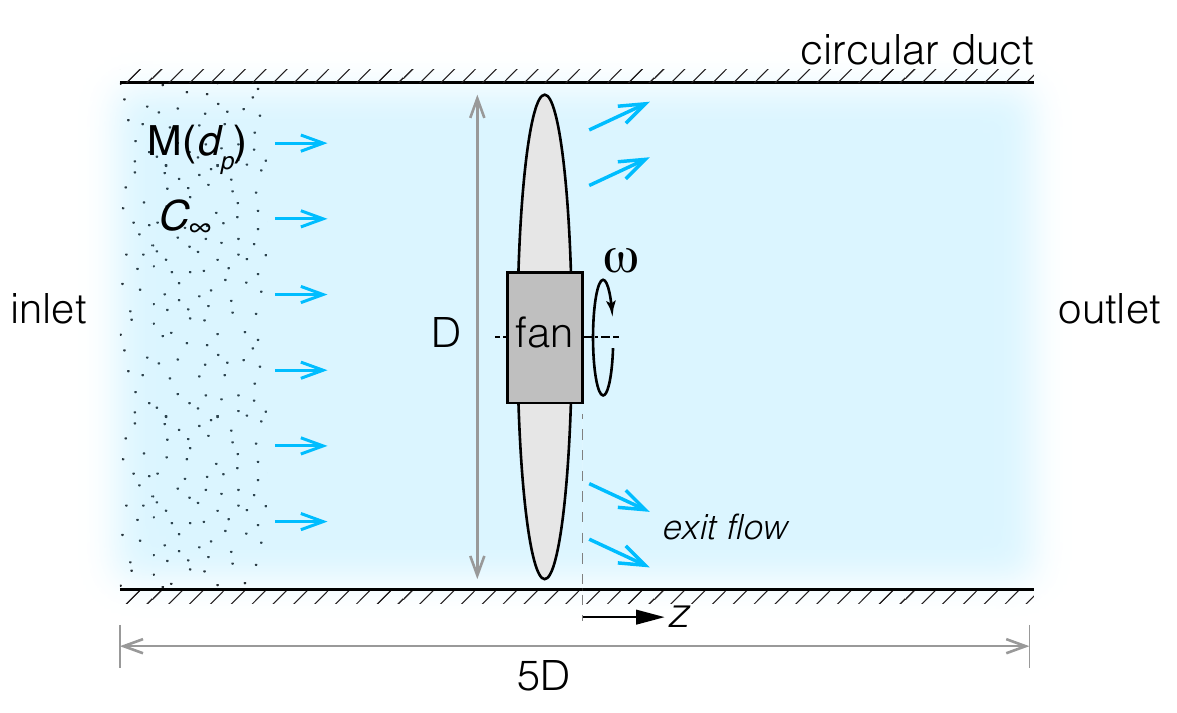}
\caption{An illustration of the computational domain.}
\label{figDomain}
\end{figure}

\noindent A seven-blade axial fan with an outer diameter of 120 mm and hub-tip ratio of 0.417 was considered for the numerical investigations. The blade profile was chosen to be a NACA 6409 aerofoil with 30 mm chord length. The angle of attack at the hub was set at 45$^{\circ}$ and the blade twist angle was -10$^{\circ}$, respectively. Fan characteristics of pressure rise and flow rate were assessed by placing the axial fan concentrically within a circular duct, providing a gap of 1.2 mm between the wall and the blade tips. The fan was located in the middle of this 600 mm long duct. An illustration of this arrangement is shown in Fig. ~\ref{figDomain}.  A constant volume flow rate was delivered at the inlet to the duct, and the pressure rise across the fan was analysed. Different flow rates spanning maximum to minimum pressure rise were applied to construct the pressure-flow characteristic and determine the recommended operating region.  

\begin{figure}[t]
\centering
\includegraphics[width=3.0in]{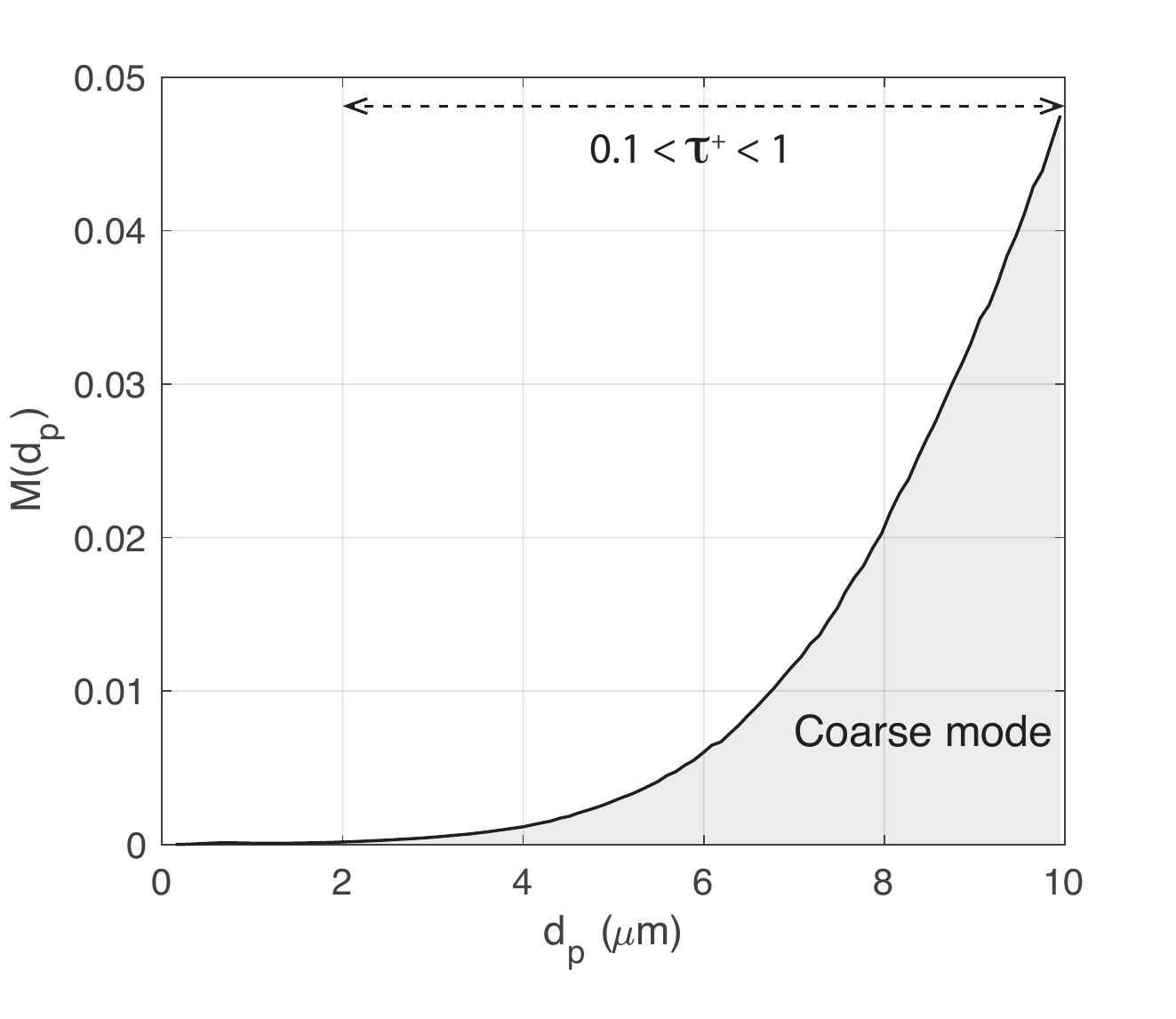}
\caption{Size distribution of the fractional mass concentration of coarse mode atmospheric particles.}
\label{figPDFs}
\end{figure}

Investigations on the transport of atmospheric pollutants under varied operating conditions were also conducted in the same configuration discussed above. A particle distribution was constructed to include the characteristic coarse mode observed in outdoor environments \cite{PMDist}. Particles in the 2.5-10 $\mu$m range are not completely removed from the environments of information technology equipment when using standard MERV 11 air filtering \cite{Stafford2021}. An indoor mass distribution of the coarse mode particles, $M(d_p)$, was created considering a cut-off at 10 $\mu$m. This resulted in a range of particle diameters shown in Fig. \ref{figPDFs}. A constant particle density of 1500 kg m$^{-3}$ was chosen based on an average of the seasonal and diurnal measurements of the apparent density of atmospheric particles \cite{Liu2015}. Particles were introduced at the inlet to the duct with an initial velocity equal to the air flow. The number of particles introduced at this boundary condition was prescribed to maintain a constant particle concentration ($C_\infty$) at the fan inlet for all operating conditions.      

\subsection{Numerical techniques}

\noindent An Eulerian-Lagrangian computational fluid dynamics approach was used to predict the transport of individual atmospheric particles in air. Transient, Reynolds-Averaged Navier-Stokes simulations (RANS) were performed using the $k-\epsilon$ turbulence model to describe the transport of turbulent kinetic energy (TKE) and the rate of dissipation of TKE. This was implemented within the open source continuum mechanics software OpenFOAM, using the PIMPLE algorithm for pressure-velocity coupling. A no-slip condition was applied at walls and the Spalding wall function was used for near-wall treatment. This Eulerian approach was coupled to a Lagrangian solver that described the particle motion according to: 
\begin{equation}
m_{i} \frac{d \mathbf{v}_{i}(t)}{d t}=\sum \mathbf{F}_{i}
\label{eq_MaxRiley}
\end{equation}

\noindent where $m_i$ is the particle mass and $\mathbf{{v}_i}$ the particle velocity. The sum of forces acting on the particle depends on the particle and carrier fluid densities \cite{Kuerten2016}. In the present work, $\rho_{i}\gg\rho_{air}$ and the dominant forces considered are particle drag ($F_D$) and gravity ($F_g$), respectively. The drag force is computed according to the Ergun-Wen-Yu equation:
\begin{equation}
\mathbf{F}_{D}=m_{i} \frac{\left(\mathbf{u}\left(\mathbf{x}_{i}(t), t\right)-\mathbf{v}_{i}\right)}{\tau_{i}}\left(1+0.15{Re}_{i}^{0.687}\right)
\label{eq_ErgunWenYu}
\end{equation}

\noindent where $\mathbf{{u}}$ is the air velocity, $\tau_{i} = \rho_{i}d_{i}^{2} / 18\mu$ is the particle relaxation time, and $Re_{i} = \rho_{air} d_{i}\left|\mathbf{u}\left(\mathbf{x}_{i}(t), t\right)-\mathbf{v}_{i}\right|/\mu$ is the particle Reynolds number.

Individual particles were modelled according to (\ref{eq_MaxRiley}) and each simulation introduced $N_{p}\sim 10^{6}$. The volume fraction of these  particle pollutants in air was $\phi_p \sim 10^{-7}$ and permitted one-way coupling between the fluid and the particle. This meant the air influenced particle motion, however, the particles had a negligible impact on the air flow properties. Turbulence effects on particles were considered using a stochastic dispersion model that obtained turbulence information from the RANS solution and applied an isotropic assumption. Particles deposited on the surface of the duct wall when they reached a particle-wall distance less than $d_p /2$. 

The motion of the axial fan was resolved directly by creating a dynamic mesh to rotate the geometry at constant angular speed, $\omega$. This approach is computationally intensive compared to compact fan models and MRF methods that are traditionally used for the thermal design of air-cooled electronic and photonic systems \cite{Stafford2014}. However, it also provided the most accurate representation of the fluid mover and supported detailed analyses on the spatio-temporal characteristics of this multi-phase flow.

\subsection{Data reduction and validation}
\noindent The rate of particle deposition onto a surface can be described by the deposition velocity:

\begin{equation}
V_{d} = \frac{J}{C_{\infty}}
\label{eq_Vd}
\end{equation}

\noindent which relates the mass flux per unit time, $J$, to the concentration of particles in the air, $C_{\infty}$. The deposition velocity depends on a number of physical processes. The typical regimes include Brownian diffusion, turbulent diffusion and eddy impaction, and inertia moderated. These are categorized through the non-dimensional deposition velocity and particle relaxation time parameters:        

\begin{equation}
V_{d}^{+} = \frac{V_{d}}{u^{*}}
\label{eq_Vd+}
\end{equation}

\begin{equation}
\tau^{+} = \frac{\tau_{p}u^{*^2}}{\nu}
\label{eq_tau+}
\end{equation}

\noindent where $u^{*}$ is the friction velocity, $\tau_{p}$ is the particle relaxation time, and $\nu$ is the kinematic viscosity. In the present work, non-dimensional particle relaxation times span the turbulent diffusion and eddy impaction regime ($0.1<\tau^{+}<1$, Fig. \ref{figPDFs}). 

\begin{figure}[t]
\centering
\includegraphics[width=3.0in]{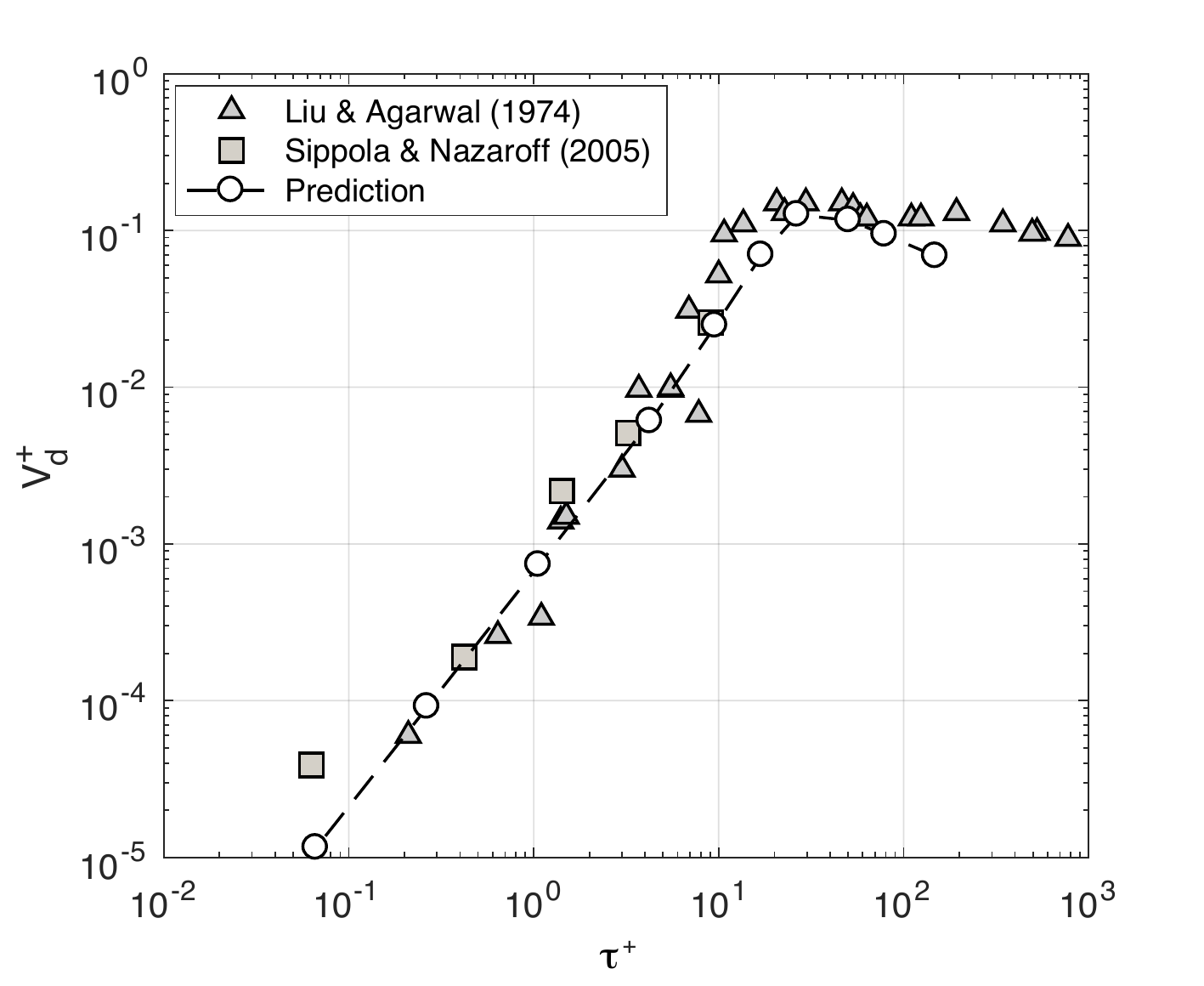}
\caption{Predicted relationship between the dimensionless particle deposition velocity, $V_{d}^{+}$, and the dimensionless particle relaxation time, $\tau{^+}$.}
\label{figFD}
\end{figure}

The numerical method was verified against experimental particle deposition data for fully developed turbulent flow in ducts \cite{Liu1974,Sippola2005}.
Fig. \ref{figFD} shows the comparisons with the predictions of the proposed model providing suitable agreement in the turbulent diffusion / eddy impaction regime and for the transition to the inertia moderated regime ($\tau^{+}\gg 1$).

\begin{table}[t]
\caption{Numerical grid sensitivity analysis}
\begin{center}
\label{tabGrid}
\begin{tabular}{c c c c}
& & \\ 
\hline
$N_c \times 10^5$ & $E(\Delta{P})$ & $E(u_x,u_y,u_z)$ & $E(m_{p,d})$ \\
\hline
3.06 & 0.064 & 0.142, 0.044, 0.202 & 0.332\\ 
4.84 & 0.060 & 0.104, 0.034, 0.165 & 0.256\\ 
8.00 & 0.037 & 0.092, 0.033, 0.136 & 0.190\\ 
13.28 & 0.015 & 0.056, 0.027, 0.112 & 0.102\\ 
22.41 & 0.021 & 0.047, 0.032, 0.097 & 0.066\\ 
41.37 & - & - & -\\ 
\hline
\end{tabular}
\end{center}
\end{table}

A sensitivity analysis was performed to determine the numerical grid parameters that ensured sufficient accuracy. Numerical solutions for six different grids were tested with the fan running at an intermediary operating point near maximum efficiency. Comparisons between the predictions for global quantities (pressure rise), local velocity profiles in three-dimensions, and total mass of particles deposited on the duct surface were used to assess the sensitivity of the grid settings on the solution accuracy. For these assessments, a non-dimensional time of $t^* = \omega{t}/2\pi = 9$ was selected across the transient simulations. These results are presented in Table ~\ref{tabGrid} where the relative differences, $E(.)$, are calculated with respect to the predictions at the finest grid resolution. Local velocity differences (RMSD) are normalized by the mean flow velocity of 3 m s$^{-1}$. The numerical grid settings for $N_c = 1.328 \times 10^6$ cells converged to a maximum difference $\approx 10\%$ and was selected to balance accuracy with computational time.

\section{Results and Discussion}

\noindent The interplay between fan operating characteristics, atmospheric particle transport, and particle deposition are discussed in this section. Following this investigation, a simple flow control solution that addresses dominant deposition mechanisms is presented for a telecommunications equipment test case.

\subsection{Airflow characteristics} \label{secPQD}

\begin{figure}[b]
\centering
\includegraphics[width=3.25in]{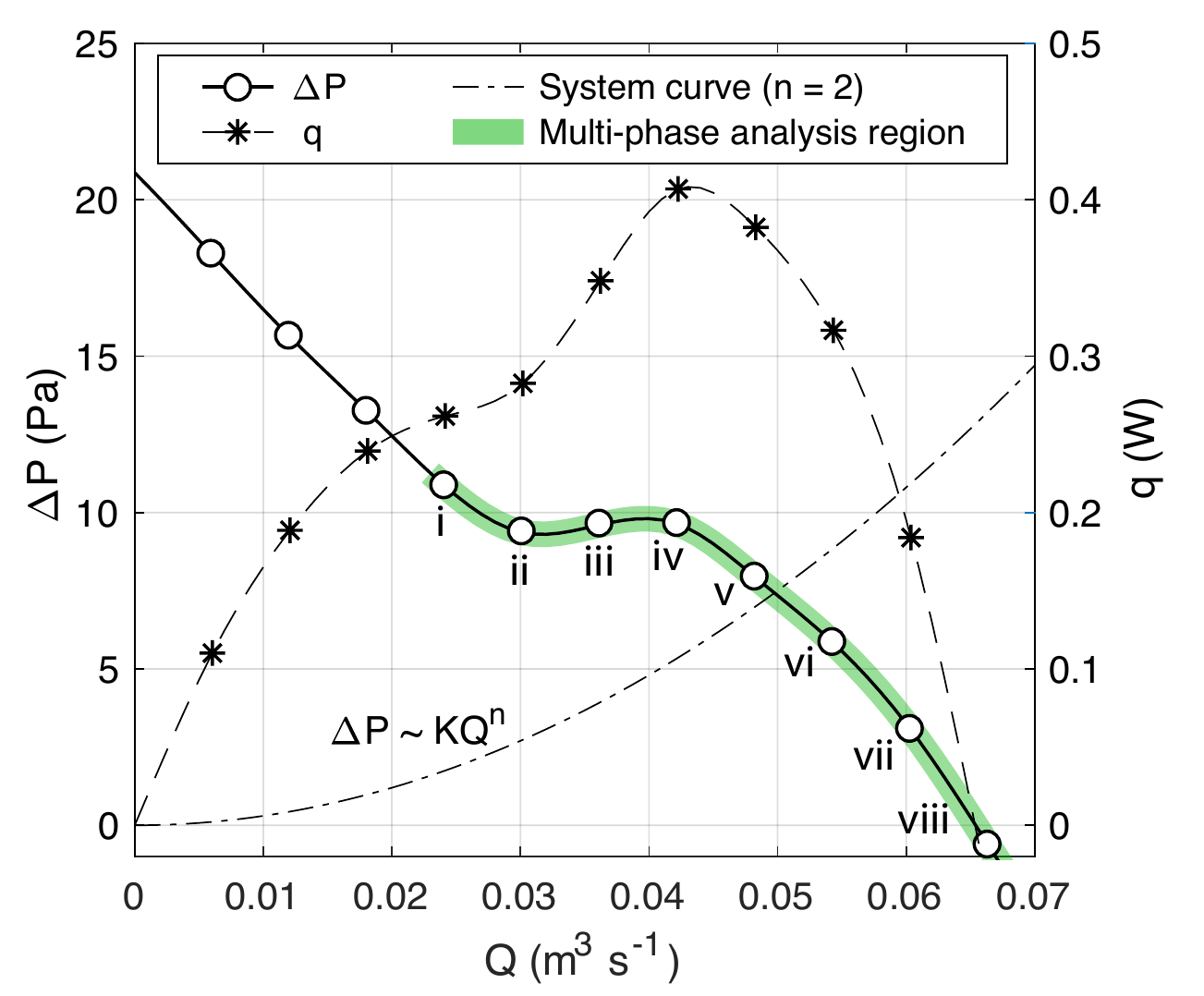}
\caption{Fan pressure rise and power curves for an operating speed of $\omega = 2000$ rpm.}
\label{figPQq}
\end{figure}

\noindent The predicted characteristic pressure-flow rate and power curves for the axial fan under investigation are shown in Fig. \ref{figPQq}. The stall region, defined by the pressure trough in the $\Delta{P}-Q$ curve (points ``i-iii"), is a regular feature in axial fan performance curves. Here, the flow is unstable and separates from the fan blades. It is a region normally avoided for cooling applications, with surging flow effects, pressure variations, propensity for increased bearing wear, vibrations and noise generation. The region toward the maximum flow rate at zero pressure rise is also typically avoided. At this location near point ``viii", the efficiency reduces considerably as reflected by the reduction in aerodynamic power, $q$. The operating region regularly designed for in fan-cooled systems is between points ``iv" and ``viii". 

The flow resistance of these systems, such as heat exchangers or printed circuit boards in Fig. \ref{figSketch}, follow a system characteristic behaviour also presented in Fig. \ref{figPQq}. This curve is a function of the flow regime (e.g. laminar/turbulent) through the exponent, $n$, and various other losses through the system accounted for by the prefactor, $K$. While most thermal designs are arrived at for a fixed system characteristic curve, there are practical scenarios where this curve can change throughout the equipment's operational lifetime. For example, filter blockages can shift the operating point towards the stall region. Similarly, the installation of additional circuit boards to card slots. Over-sizing or under-sizing fans for cooling applications can also result in adverse operating conditions. To capture this range of possible operating scenarios, a wide exploration space has been considered for multi-phase flow simulations. This is highlighted in Fig. \ref{figPQq} and spanned from stall to free delivery conditions. 

\subsection{Particle transport and deposition mechanisms} \label{secMechanisms}
\begin{figure}[b]
\centering
\includegraphics[width=3.25in]{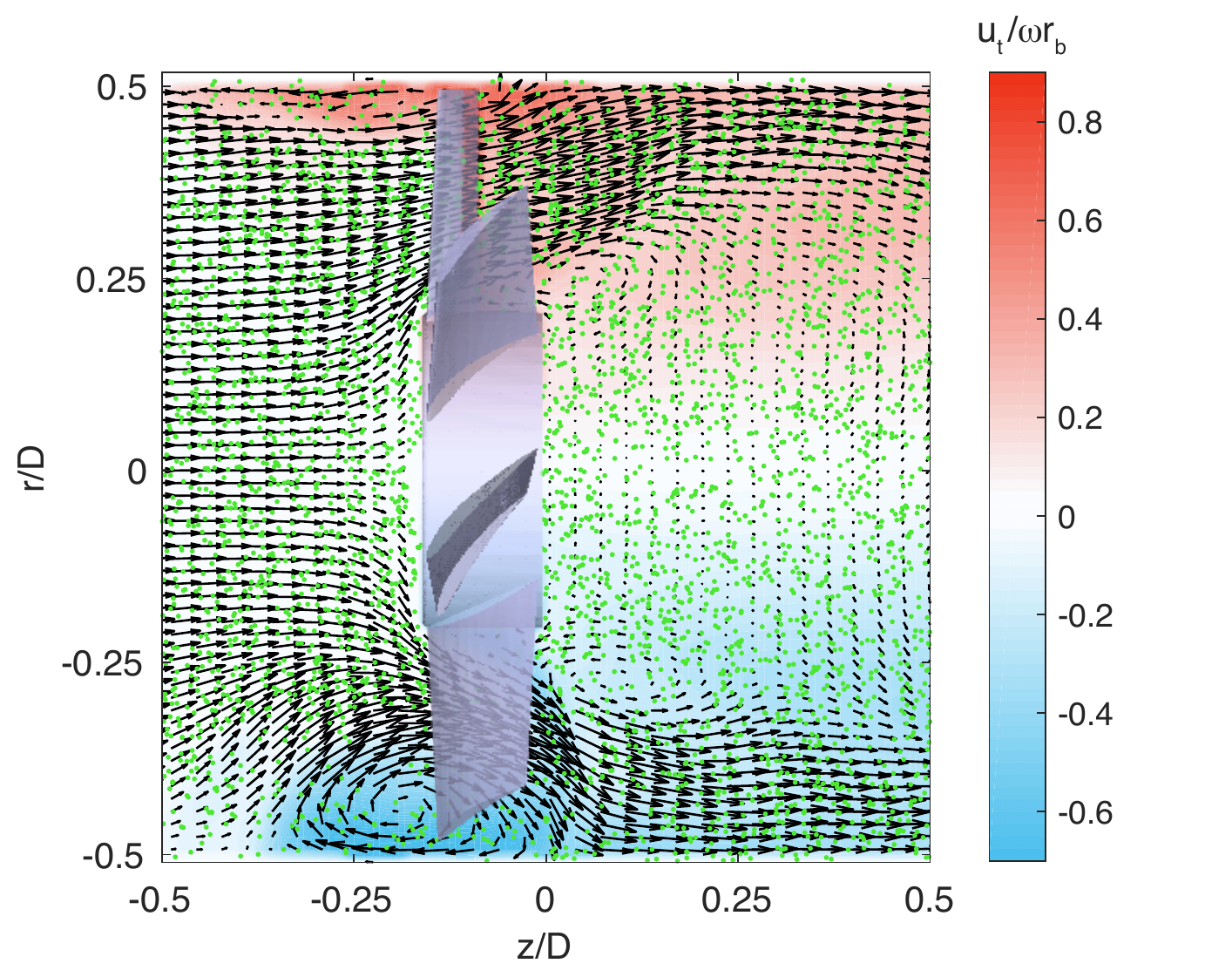}
\caption{A snapshot of the three-dimensional air and particulate flows at $\omega = 2000$ rpm and $t^{*} = 10$. Particles have been colored green. Airflow is represented by velocity vectors in the radial-axial plane and contours of tangential velocity normalized by blade tip speed.}
\label{fig3D-snapshot}
\end{figure}

\noindent Enhanced particle fouling in forced-cooled systems, illustrated in Fig. \ref{figSketch}, predominantly occurs in close proximity to the fan exit where the highest shear velocity and turbulent stresses exist \cite{Stafford2021}. Therefore, an area $\pm 0.5(z/D)$ upstream and downstream of the fan exit flow was the focus for this study. Fig. \ref{fig3D-snapshot} shows a snapshot of the flow field in this region for a severely stalled operating condition (point ``i", Fig. \ref{figPQq}). This two-dimensional plane intersects a fan blade at mid-chord for positive $r/D$. In the lower half of the plot, at negative $r/D$, the plane is located at the intermediate angle between two adjacent fan blades. This plane was selected as it captures both the air flow during blade passage (+ve $r/D$) and in the blade wake (-ve $r/D$).

Two dominant flow features are observed. The first are the large vortices which emanate from the blade tip region, severely distorting the inlet flow to the fan. This flow recirculation at the inlet also influences the exit flow that continues downstream. Air and particles impinge on the duct wall at an off-axial exit angle with a significant radial and tangential velocity component. This flow behaviour is promoted by a reduction in axial velocity in the stall region, the consequences of a reduced flow rate at this operating point.

\begin{figure}[t]
\centering
\includegraphics[width=3.25in]{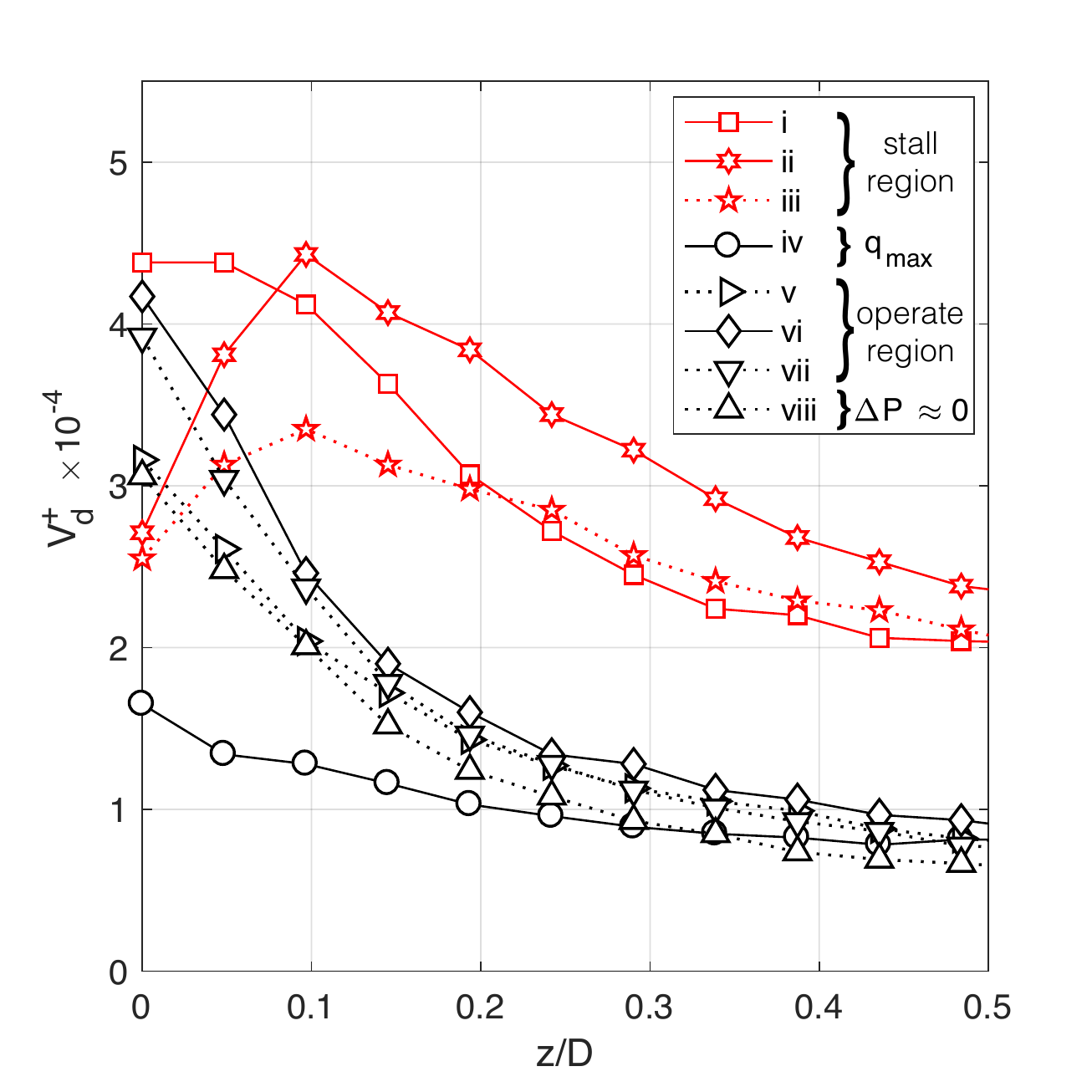}
\caption{Axial profiles of the non-dimensional deposition velocity for each operating point and a rotational speed, $\omega = 2000$ rpm.}
\label{figVdPlus}
\end{figure}

The effect of these exit flow features on the deposition characteristics have been examined from $0 \leq z/D \leq 0.5$. Fig. \ref{figVdPlus} presents the non-dimensional deposition velocities for all operating points investigated across the fan performance characteristic. These deposition data for the circular duct have been circumferentially averaged. Distinctive differences in deposition profiles were found between recommended operation and when the fan enters stall conditions. Local non-dimensional deposition velocities across $0 \leq z/D \leq 0.5$, are up to four times higher during stall operation compared to the lowest deposition case. This case occurred at the maximum fan aerodynamic power, $q_{max}$. The shape of these profiles also shifts from a monotonically decreasing deposition velocity with increasing $z/D$, to a non-monotonic behaviour as stall is established.  

These observations have important consequences for the practical design of fan-cooled electronic systems. Firstly, the mass flux of coarse mode particles with $d_{p} <$ 10 $\mu$m has significant local variations over a relatively short distance downstream. In the operating region, the non-dimensional deposition velocities span $V_{d}^{+} \sim 10^{-3}-10^{-4}$. This sharp reduction in deposition velocity agrees with the locations of patterns of preferential deposition observed on equipment operating in the field \cite{Stafford2021, Shah2019}. Furthermore, if operating outside of the recommended aerodynamic conditions, positioning sensitive components further downstream of the exit flow does not necessarily correlate with a reduced exposure to particle deposition. 

Secondly, a significant change in particle deposition can occur with small changes in operating point. For example, deposition velocity varies by a factor of 3-4 when moving $\pm 10\%$ either side of the maximum fan power. In many cooling and ventilation systems, the target fan operation point is within $15\%$ of the maximum efficiency, thus placing it in a region where high variability in particle mass flux occurs. Therefore, physical designs for forced air cooling in poor air quality indoor and outdoor environments should carefully consider holistically the fan operation, thermal design and pollution exposure levels to avoid the failure modes associated with particle deposition (e.g. corrosion, mechanical, thermal).      

\begin{figure}[t]
\centering
\includegraphics[width=3.25in]{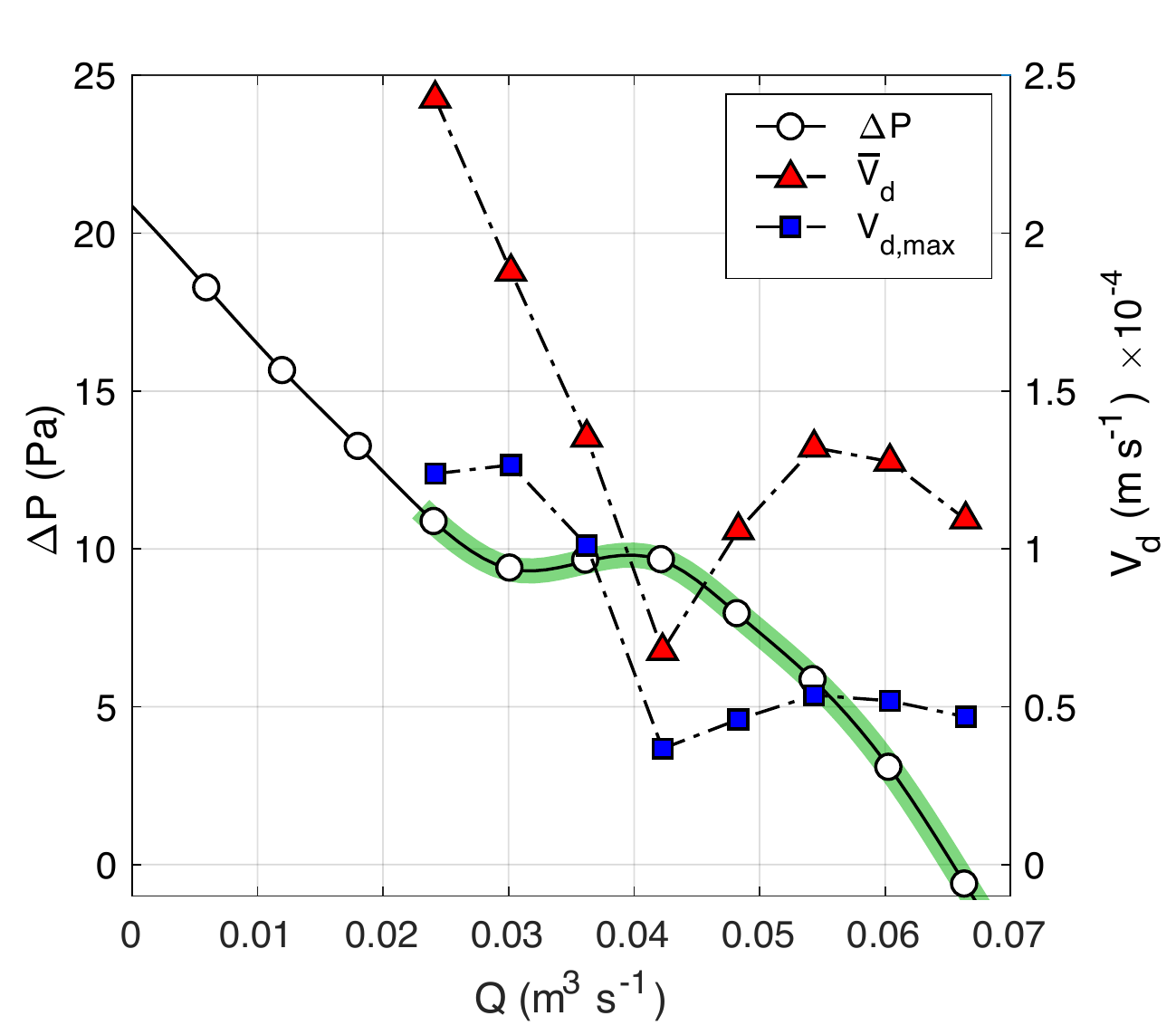}
\caption{Particle deposition behaviour operating along the fan characteristic curve for a rotational speed, $\omega = 2000$ rpm.}
\label{figPQD}
\end{figure}

These trends in deposition with operating point become more apparent by overlaying particle deposition curves on the fan performance characteristic. Fig. \ref{figPQD} shows two curves for average and maximum deposition velocity, $V_{d}$, over the region of interest, $0 \leq z/D \leq 0.5$. As noted above, the deposition minima occur at the point of maximum fan power (point ``iv" in Fig. \ref{figPQq}), with sharp increases either side of this. On the stall side, deposition velocity rapidly increases as the flow becomes unstable. Within the recommended operating region, deposition velocity increases at a lower rate, reaching a maximum mid-way between maximum fan power ($q_{max}$) and the free delivery condition ($\Delta{P} \approx 0$). 

Notably, at this operating point, there is a twofold increase in average deposition velocity compared to $\overline{V}_{d}$ at maximum fan power. The presence of this local maximum within the recommended fan operation region also opens up the opportunity for multi-objective fan and thermal design which incorporates mass transport. The weighting of objectives in this design framework (e.g. maximize heat dissipation, maximize fan efficiency, and minimize particle deposition) might depend on the application, operating environment ($T_{\infty}$, RH) and air quality conditions ($C_{\infty}$, PM$_{2.5}$, PM$_{10}$).       

The monotonic non-dimensional deposition velocity profiles across the operating region (points ``iv" to ``viii") are $V_{d}^{+} \sim 10^{-3}$ for $z/D \lessapprox 0.25$. Further downstream, for $z/D \gtrapprox 0.4$, this reduces to $V_{d}^{+} \sim 10^{-4}$ as the turbulent diffusion and eddy impaction deposition mechanisms weaken. In contrast, the unstable flows generated as the fan experiences stall remain at $V_{d}^{+} \sim 10^{-3}$ over the entire $z/D$ examined. The transition to these non-monotonic deposition profiles while operating in the stall region can be explained by the accompanying alteration to the air and particulate flow behaviour.

\begin{figure*}[t]
\centering
\includegraphics[width=7in]{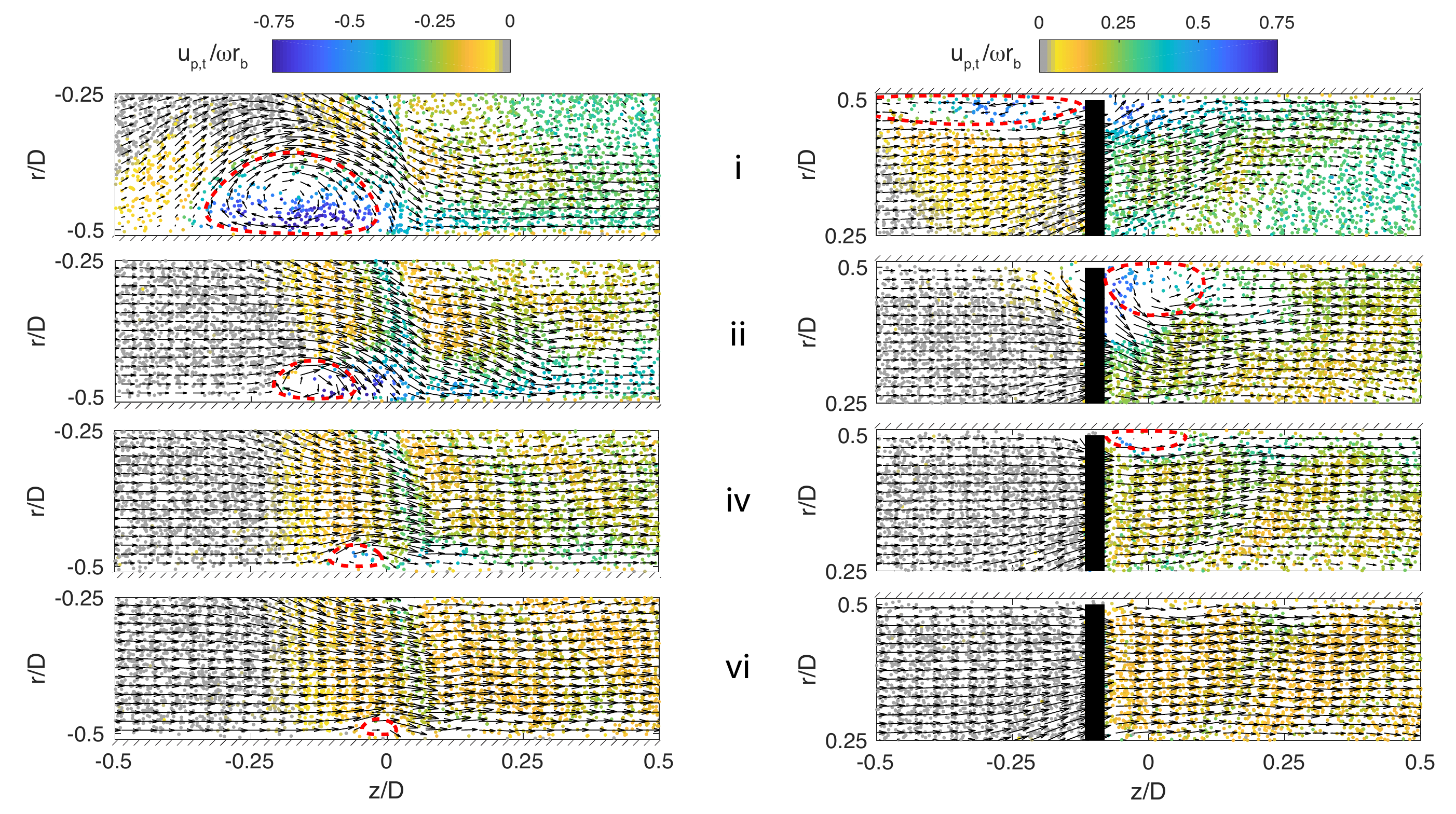}
\caption{The flow of microscale particles in the blade wake and mid-chord regions (left and right) for $\omega = 2000$ rpm and $t^{*} = 10$. Operating points correspond with stall conditions (``i" and ``ii"), maximum aerodynamic power (``iv"), and peak deposition velocity in the recommended region of operation (``vi"). Duct wall, fan inlet and fan exit flow planes are located at $r/D = \pm 0.51$, $z/D = -0.16$ and $z/D = 0$, respectively.}
\label{figDispersion}
\end{figure*}

Close-up views of the air velocity vectors and particle flow in the blade wake ($-0.5 \leq r/D \leq -0.25$) and blade mid-chord planes ($0.25 \leq r/D \leq 0.5$) for four sample fan operating points (``i",``ii",``iv" and ``vi") are shown in Fig. \ref{figDispersion}. Fan rotation introduces three dimensional flow behaviour with an increase in out-of-plane tangential particle speed, $\left|u_{p,t}\right|$, as the operating point is shifted towards higher pressure rise. Additionally, vortices form around the blade and grow in size as pressure rise increases. As discussed previously, in the severe stall case (``i") these vortices extend into the inlet side and alter the inflow velocity distribution. Interestingly, the flow maldistribution along the blade span produces protective low concentration bubbles, with many particles diverted around the vortex. Particles feed into the oblique jet that is formed at the fan exit in the stalled condition, resulting in high momentum particle impingement at the duct wall for $0 < z/D < 0.15$. This leads to a local enhancement of the wall normal mass flux, $J$, reflected in the deposition velocity profile in Fig. \ref{figVdPlus}.

As the operating point is shifted to the trough of the stall region (``ii"), the wake vortex reduces in size while the blade vortex flips to the pressure side of the fan during blade passage. The latter produces a low concentration bubble spanning $0 < z/D < 0.1$. This flow feature explains the reduction in deposition velocity for points ``ii" and ``iii" in Fig. \ref{figVdPlus}, followed by a recovery and peak in $V_{d}^{+}$ downstream as particles impinge onto the duct wall.                 

At maximum aerodynamic power (``iv"), the lowest local, maximum and average deposition velocities were observed in Figs. \ref{figVdPlus} and \ref{figPQD}. The factors contributing to reduced mass transport include: 1) it is the lowest flow rate with stable operation; 2) the tip vortex and low concentration bubble is retained in the near wall region; and 3) the radial velocity component in the exit flow decreases, reducing flow impingement and lowering the wall normal mass flux. The latter is a feature dependent on pressure rise, with the impingement angle changing from approximately $-56^\circ$ (``i"), $-47^\circ$ (``ii"), $-22^\circ$ (``iv") to $-18^\circ$ (``vi") along the fan characteristic. Collectively, these three factors reduced the local particle concentration and shear velocity immediately downstream. 

The peak deposition velocities in the operating region, $\overline{V}_{d}$ and $V_{d,max}$, occurred at point ``vi", mid-way between $q_{max}$ and $\Delta{P} \approx 0$. Here, the blade tip vortex observed in other cases no longer exists on the blade pressure side, leading to enhanced mass transport to the wall at $r/D \approx 0.5$. The vortex persists in the blade wake at $z/D < 0$, however, it is much reduced in scale compared to the previous operating points. The flow exits and impinges at $r/D \approx -0.5$ at an off-axial, oblique angle which promotes mass transfer for $0 < z/D < 0.1$ above other operating points. This oblique exit flow angle reduces as pressure rise across the fan is lowered further, reducing the mass transfer enhancements driven by this impinging flow. This results in progressively lower deposition velocities for operating points ``vii" and ``viii" at free delivery, offsetting mass transfer enhancement from an increase in flow rate.

\subsection{A flow control solution for installed equipment}

\begin{figure*}[htbp]
\centering
\includegraphics[width=6.85in]{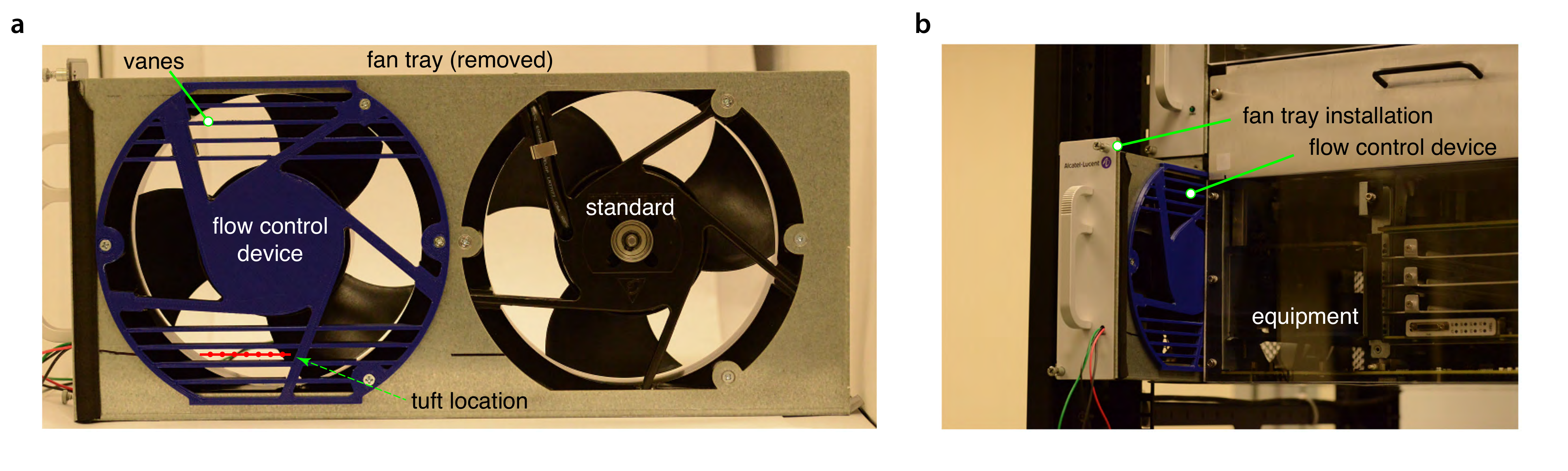}
\caption{A simple flow control solution for reducing particle impingement effects. a) A 3D printed flow control device mounted to the left fan housing within a fan tray. b) Fan tray installation in a telecommunications base station equipment.}
\label{figFanTray}
\end{figure*}

\begin{figure*}[htbp]
\centering
\includegraphics[width=4in]{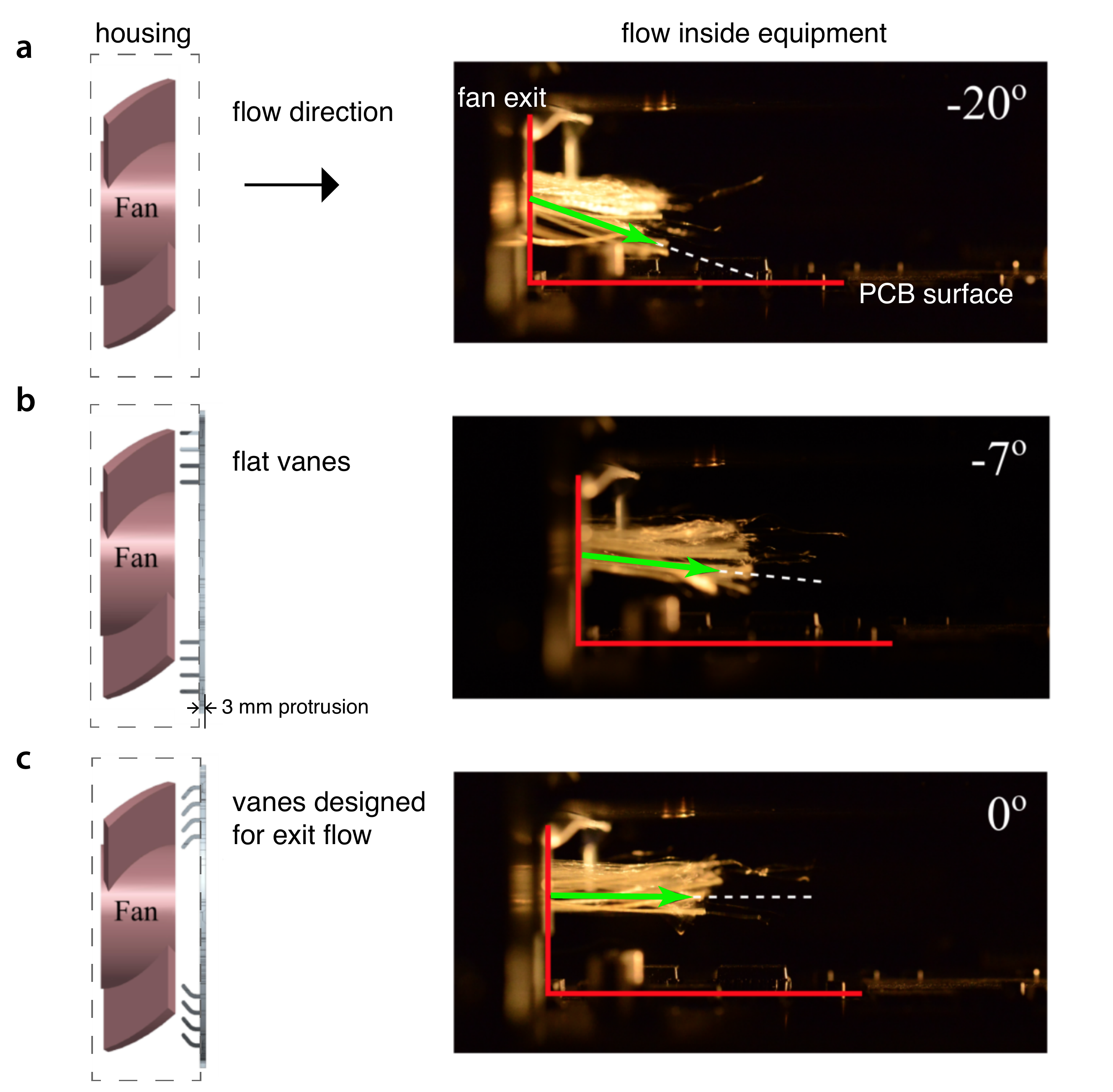}
\caption{The flow control solution (left) alters exit flow impingement within the telecommunications equipment as represented by changes in fibre tuft angle (right). The images demonstrate exit flow for a) standard operation, b) control using flat vanes, and c) control using vanes designed for the exit flow.}
\label{figTufts}
\end{figure*}

\noindent The multi-phase numerical methods and data analyses presented in this work can be used as a predictive tool for designing reliable forced-cooled electronic equipment in poor air quality environments. For equipment already in service, however, the opportunity to modify board layouts, fan designs, and other corrective hardware maintenance to solve reliability issues is generally not feasible and cost-prohibitive. An alternative approach has been considered that performs remediation on already-deployed equipment using a basic flow control device \cite{IPCpaper}.

Qualitatively, the particle transport and deposition mechanisms uncovered in this numerical investigation are ubiquitous for axial fans and recently confirmed in experiments by the authors using particle image velocimetry \cite{Stafford2021}. The investigations on the impact of operating point in Section \ref{secMechanisms} have demonstrated that fan exit flow angle has a dominant contribution to particle deposition velocity.

Using telecommunications base station equipment as the test case, shown in Figs. \ref{figFanTray} and \ref{figTufts}, stationary aerodynamic guide vanes were designed and installed on individual fans within the fan tray. These altered the exiting air flow angle so that the oblique flow impingement was reduced, while limiting adverse effects to the bulk aerodynamic performance of the fan ($< 5\%$ reduction in volumetric flow rate). This concept utilized the space available in the fan housing and protruded by only 3 mm outside of this, allowing for installation of the existing fan tray without any equipment modifications. 

For the purposes of the study, and to investigate the alteration to exit flow angle, the equipment was modified with a transparent face plate to permit flow visualization using fibre tufts. These tufts were placed in the exit flow plane at locations indicated in Fig. \ref{figFanTray}a and images were captured using a DSLR camera and halogen light source during fan operation. The vanes designed to have a flow entrance angle matching the exit flow in Fig. \ref{figTufts}a, and a flow exit angle aligned parallel to the printed circuit board, altered the downstream flow angle from $-20^{\circ}$ to $0^{\circ}$. This concept, therefore, provides a simple, low cost solution to one of the main drivers for particle deposition.

\section{Conclusion}
\noindent Indoor and outdoor electronic equipment located in poor air quality environments presents reliability challenges. Forced cooling, although integral in the prevention of thermally-induced failures, indirectly exposes components and devices to particle pollutants. In this study, multi-phase numerical simulations on axial fan flows were performed to determine the influence of operation point on the deposition of coarse mode atmospheric particles with diameters below 10 $\mu$m. Deposition spanned the turbulent diffusion and eddy impaction regime and the dominant deposition factors across the fan characteristic, from aerodynamic stall to free delivery, have been revealed. Deposition velocity was highest in the stall region, with particle momentum and wall normal flux enhanced by the instabilities and oblique flow impingement near the fan exit. Deposition velocity was lowest at maximum aerodynamic power, as the presence of blade tip vortices and a maldistribution of flow along the blade span provided a protective bubble of low particle concentration, limiting the local mass transport of particles to the surface. Within the normal operation region, a local peak in deposition velocity was found, as a trade-off between increased flow rate and reduced impingement angle at low pressure rise occurred. Finally, a practical, low cost flow control solution was developed and installed into real telecommunications equipment to alter particle impingement angle and reduce deposition rate.


%



\section*{Acknowledgment}
\noindent J.S. acknowledges use of High Performance Computing facilities at the University of Birmingham (BlueBEAR). J.S. would also like to acknowledge Mr. Jahan Hadidimoud who provided the fan CAD geometry for the numerical simulations performed in this study.

\vspace{-1cm}
\begin{IEEEbiography}{Jason Stafford} is a lecturer in the School of Engineering at the University of Birmingham. He earned his Ph.D. in experimental thermofluids at Stokes Institute, University of Limerick, Ireland. He worked at Bell Labs as a Member of Technical Staff and joined Imperial College London in 2017 as a Marie Skłodowska-Curie Fellow. His research interests are at the interface between fluid dynamics, heat and mass transport phenomena, and materials science, with application to electronics thermal control and solution processing of nanomaterials.
\end{IEEEbiography}

\vspace{-1cm}
\begin{IEEEbiography}{Chen Xu}
Biography text here.
\end{IEEEbiography}





\end{document}